# INHOMOGENEOUS MAGNETIC FIELD INFLUENCE ON MAGNETIC PROPERTIES OF NiFe/IrMn THIN FILM STRUCTURES

Ch. Gritsenko[1], A. Omelyanchik[1], A. Berg[1], I. Dzhun[2], N. Chechenin[2,3], O. Dikaya[1], O. A. Tretiakov[4,5], V. Rodionova[1,6]

[1]*Immanuel Kant Baltic Federal University, A. Nevskogo 14, 236041 Kaliningrad, Russia*
[2]*Skobeltsyn Institute of Nuclear Physics, Lomonosov Moscow State University, Leninskie Gory 1/2, 119991 Moscow, Russia*
[3]*Faculty of Physics, Lomonosov Moscow State University, Leninskie Gory 1/2, 119991 Moscow, Russia*
[4]*Institute for Materials Research and Center for Science and Innovation in Spintronics, Tohoku University, Sendai 980-8577, Japan*
[5]*School of Physics, The University of New South Wales, Sydney 2052, Australia*
[6]*National University of Science and Technology MISIS, Leninsky Prospect 4, Moscow, 119049, Russia*

Abstract
We demonstrate how the configuration and magnitude of a magnetic field, applied during magnetron sputtering of a NiFe/IrMn bilayer, influence the magnetic properties of the structure, such as hysteresis loop shape, coercivity, and exchange bias. Furthermore, we illustrate that it is possible to create a stepwise hysteresis loop in the sample's region with the highest field gradient. The found features can be used for future sensor applications.

PACS: 75.80.+q, 96.12.Hg, 75.50.Bb

Keywords: NiFe/IrMn structure, inhomogeneous applied field, magnetization process; magnetic anisotropy

## 1. Introduction

A large exchange bias effect in ultrathin or granular structures is employed for sensors and memory applications [1–3]. Large thickness of ferromagnetic layers and existence of exchange bias, independent of its value, are important factors for sensors based on the magnetoimpedance effect [4–6]. Exchange bias in polycrystalline films is employed in read-head technology and MRAM devices due to the possibility of tuning magnetic properties of ferromagnets (FMs) and antiferromagnets (AFMs) through the proximity effects in FM/AFM thin film structure [2, 7]. Another application, requiring tunable magnetic properties by means of an in-plane magnetic anisotropy of soft magnetic thin films is monolithic microwave integrated circuits [8]. This in-plane magnetic anisotropy can be induced by an exchange coupling with an AFM layer. Magnetic features of exchange-biased systems can be used to create stepwise magnetic hysteresis loops, which have a potential to increase the sensitivity and implement the secure passive magnetic tags [9–11].

In this application, a rich harmonic spectrum of signal induced in pick-up coils is required. The features in the spectrum can be caused by the magnetization reversal processes of magnetic systems with peculiar and easily tunable hysteresis loops.

The tuning of magnetic properties (hysteresis loops shape, coercivity, and exchange bias) of thin-film structures with exchange bias for different applications can be realized by changing either the deposition conditions or the method of sample synthesis [7, 12, 13]. To observe a controllable stepwise hysteresis loop for an exchange-coupled structure, it should typically comprise of at least two ferromagnetic layers [14–16]. Two FM layers can possess either both in-plane or out-of-plane anisotropies, or one can have in-plane and the other out-of-plane anisotropy [17–20].

Depending on the composition of the ferromagnetic layer and sputtering conditions, growth of large grains in layers can occur, and thus the exchanged-biased bilayers may exhibit asymmetrically-shaped hysteresis loops owing to an in-plane reorientation of the ferromagnet easy magnetization axis [21–23]. To obtain exchange bias, the sample of exchange-coupled systems should be grown in the presence of a homogenous magnetic field, sufficiently strong to induce an unidirectional anisotropy.

In this work we consider a novel method for achieving a controllable stepwise hysteresis loops by applying an inhomogeneous magnetic field during deposition of a FM/AFM structure using a magnetron sputtering system.

## 2. Experimental details

The main idea of this study is to control magnetic properties of FM/AFM structures using an inhomogeneous magnetic field that was applied during thin film deposition. For this purpose, we have used two NdFeB permanent magnets of curved shape to create an external magnetic field in the deposition camera of a magnetron sputtering tool. The (100) oriented Si substrate (10*20 mm$^2$) was placed between these two magnets such that it remained stationary relative to the magnets during the deposition process, with the magnetic field being applied in-plane to the substrate. Thus, various magnitudes and configurations of magnetic field were realised in the substrate plane (Fig. 1).

Figure 1 illustrates the distribution of the magnetic field simulated by *Comsol Multiphysics*. The arrows show the direction of the magnetic field. The black frame schematically shows the region where the substrate was placed. Thus, one can observe that the substrate is located in regions with either high curvature of magnetic field or almost homogeneous magnetic field.

The magnitude of the magnetic field between the two magnets was mapped point by point using a Hall probe, with a step of 2 mm. The measured values are presented in Fig. 2. The black grid shows the locations of substrate regions. One can therefore conclude that the regions 3, 4, 5, and 6 are located in the most homogeneous magnetic field, whereas regions 1, 2, 7, and 8 are in the most inhomogeneous field.

The deposition process was thus performed in the presence of in-plane magnetic field with the above described magnitude and configuration. The sputtering process was done in an Ar atmosphere with a pressure of $3*10^{-3}$ Torr. The layer thicknesses were set by the deposition time. The deposition rates were estimated from measurements of the thickness of the calibration samples using Rutherford backscattering.

Thin film structures were deposited onto a (100) Si substrate. The buffer Ta layer with a thickness of 30 nm was deposited onto the substrate to improve the growth of further layers. Then, 10 nm of $Ni_{75}Fe_{25}$ layer was fabricated by co-deposition from the two separate targets. After that, the layer of $Ir_{45}Mn_{55}$ with a thickness of 20 nm was sputtered from the single alloy target. The last layer of 30 nm of Ta was deposited on top of all structures to prevent them from oxidation. After the deposition, the thin film was cut into pieces with the area of 5*5 mm$^2$.

The crystalline structure of samples was studied with an X-ray diffractometer (Bruker D8 Discover) using Cu radiation with wavelength of 0.154 nm. Figure 3 shows typical XRD patterns of these structures, with clearly observable peaks for Ta (200), IrMn (111), and NiFe (111) crystallographic planes. Such peaks are usually observed for exchange-biased NiFe/IrMn systems [24, 25]. Atomic Force Microscopy (AFM) imaging showed that the roughness of individual NiFe and IrMn films, grown on Ta buffer, is of 2 nm and 1 nm, respectively [26]. Therefore, we assume the interface between the NiFe and IrMn to be rather smooth.

The magnetic properties of the samples were investigated using a Vibrating Samples Magnetometer (VSM, Lake Shore, Model 7400). For each sample, the hysteresis loops were obtained for in-plane geometry of the magnetic field of the VSM at different azimuthal angles. The azimuthal angle of 0° corresponds to the co-directed field of VSM and applied field during the deposition marked as y axis in Fig. 1 (i.e., perpendicular to the surfaces of the magnetic pole plates).

### 3. Results and discussion

The reasoning provided below is based on observing the inhomogeneity of the magnetic field described in Sec. 2 (Figs. 1 and 2). One can note that since the magnetic field applied during sample deposition in regions 3 and 4 was found to be uniform, the azimuthal dependence of exchange bias for the samples from these regions shows classical behavior for exchange-coupled bilayer structures with in-plane unidirectional anisotropy. The maximal exchange bias and rectangular hysteresis loop is detected along the easy axis of magnetization of the film (angles 0°, 180°, 360°, See Fig. 4), whereas the exchange bias being absent, coercivity being near zero and hysteresis loop being tilted along the hard axis of magnetization (angles 90° and 270°). An existence of increased exchange bias at the angles multiple of 45° can be explained by different effects, for example, by the occurrence of a domain wall in AFM-layer [27, 28]. But this needs further investigations.

For regions 5 and 6, the partial deviation of the magnetic field direction from that of regions 3 and 4 is observed. Thus, the corresponding deviation of the uniaxial anisotropy in small areas in the plane of the film should appear because of a distribution of the local anisotropy (Figure 5 is a scheme illustrating this effect). As a result, the exchange bias along the easy axis of magnetization is slightly less than corresponding values for regions 3 and 4, where the uniaxial anisotropy is stronger, and the exchange bias of small values up to 9 Oe arises along the hard axis of magnetization.

For regions 1 and 2, there is observable increased inhomogeneity of the magnetic field with significant deviation of the field lines from those in regions 3 and 4, therefore the unusual exchange bias behavior for the samples from these two regions is found: (i) the large exchange bias accompanied by the large coercivity at additional angles of 45° (negative shift) and 225° (positive shift), (ii) tilting of the hysteresis loops measured at 0° and 180° near switching field, and (iii) larger values of the exchange bias at 90° and 270° than those for previous regions. Hence, in these regions there is a co-existence of two areas with different orientation of a unidirectional anisotropy axis (schematically shown in Fig. 5). Moreover, the hysteresis loops taken at angles 45°, 315° and 135°, 225° coincide in pairs for regions 3 and 4, but not for regions 1 and 2. This fact, in addition to (i) and (ii), means that the declination of a unidirectional anisotropy axis in parts of the sample has angles less than 90° and the anisotropy energy of the area with more declined axis is quite small because an inhomogeneity of magnetic field in the region leads to distribution of local anisotropy as it was found for region 5 and 6.

The magnetic field in regions 7 and 8 is the most twisted compared with all previous regions (Fig. 1). As a result, the hysteresis loops of the samples from these regions are of the highest interest because they manifest the presence of a stepwise shape loop. There are observed full separations of hysteresis loops into two sub-loops similar to trilayer thin-film structures with exchange bias [29, 30]. The hysteresis loops for region 7 are shown in Figure 6. Thus, we can conclude that the planes of the films are parted into two almost identical areas (Fig. 5) with magnetizations $M_1$ and $M_2$, which are determined by the step heights of the hysteresis loops. Regarding the sub-loops being located on both sides of the Y-axis, we can suggest that these two areas have opposite orientation of the unidirectional anisotropy axis. The complex forms of hysteresis loops appear during magnetization reversal at all angles. The angular dependence of hysteresis loops for samples from regions 7 and 8 is similar to that for regions 1 and 2.

The angular dependences of the exchange bias and coercive force, estimated from the hysteresis loops, are presented in Figure 7. For regions 7 and 8 the values were estimated separately for every sub-loop, i.e. for each area. From these dependencies, one can elucidate the difference in anisotropy properties of the samples discussed in this section.

### Conclusions

We have demonstrated that presence of an inhomogeneous magnetic field during the deposition leads to dramatic changes in the magnetization reversal process of an exchanged-coupled thin film structures. A low gradient of the magnetic field results in altered values of the exchange bias. Meanwhile, a large gradient affects both the magnitude of the exchange bias and the magnetization reversal mechanism of the NiFe/IrMn bilayer thin film. These features provide the ability to expand the field of applications of the exchange bias effect.

**Acknowledgments**

O.A.T. acknowledges support by the Grants-in-Aid for Scientific Research (Grant Nos. 17K05511 and 17H05173) from MEXT, Japan, by the grant of the Center for Science and Innovation in Spintronics (Core Research Cluster), Tohoku University, and by JSPS and RFBR under the Japan-Russian Research Cooperative Program. We thank Dr. Mikhail Gorshenkov for helpful discussions.

**References**

[1] K. Watanabe, Y. Miyamoto, K. Nishimura, S. Nakagawa, M. Naoe, Investigation of the Exchange Coupling Field between NislFe19/FesoMnso Bilayers in Spin Valve Devices by Ion Bombardment to Interfaces, J. Magn. Soc. Japan. 21 (1997).

[2] C. Tannous, R.L. Comstock, Springer Handbook of Electronic and Photonic Materials, 2017. doi:10.1007/978-3-319-48933-9.

[3] T. Yu, X.K. Ning, W. Liu, J.N. Feng, D. Kim, C.J. Choi, Z.D. Zhang, Exchange coupling in ferromagnetic/antiferromagnetic/ferromagnetic [Pd/Co]n/NiO/Co trilayers with different [Pd/Co] anisotropy, J. Magn. Magn. Mater. 385 (2015) 230–235. doi:10.1016/j.jmmm.2015.03.028.

[4] V.. Vas'kovskiy, V.. Lepalovskij, A.N. Gor'kovenko, N.A. Kulesh, S. P.A., A.V. Svalov, S. E.A., N.N. Shcheglova, A.A. Yuvchenko, Magnetoresistive medium based on the film structure of Fe20Ni80 / Fe50Mn50, J. Tech. Phys. 85 (2015) 118–125.

[5] C. García, J.M. Florez, P. Vargas, C.A. Ross, Asymmetrical giant magnetoimpedance in exchange-biased NiFe, Appl. Phys. Lett. 96 (2010) 18–20. doi:10.1063/1.3446894.

[6] G. V Kurlyandskaya, A. García-Arribas, E. Fernández, A. V Svalov, Nanostructured Magnetoimpedance Multilayers, IEEE Trans. Magn. 48 (2012). doi:10.1109/TMAG.2011.2171330.

[7] K. O'Grady, L.E. Fernandez-Outon, G. Vallejo-Fernandez, A new paradigm for exchange bias in polycrystalline thin films, J. Magn. Magn. Mater. 322 (2010) 883–899. doi:10.1016/j.jmmm.2009.12.011.

[8] W. Xu, J. Zhu, S. Member, Y. Zhang, Y. Guo, G. Lei, New Axial Laminated-Structure Flux Switching Permanent Magnet Machine with 6/7 Poles, IEEE Trans. Magn. 47 (2011) 2823–2826.

[9] V. Rodionova, M. Ilyn, M. Ipatov, V. Zhukova, N. Perov, N Perov, A. Zhukov, Spectral properties of electromotive force induced by periodic magnetization reversal of arrays of coupled magnetic glass-covered microwires, Addit. Inf. J. Appl. Phys. V C. 111 (2012) 7–706. doi:10.1063/1.3680529doi:10.1063/1.3680529.

[10] R.R. Fletcher, N.A. Gershenfeld, Remotely interrogated temperature sensors based on magnetic materials, IEEE Trans. Magn. 36 (2000) 2794–2795. doi:10.1109/20.908592.

[11] K.G. Ong, D.M. Grimes, C.A. Grimes, Higher-order harmonics of a magnetically soft sensor: Application to remote query temperature measurement, Appl. Phys. Lett. 80 (2002) 3856–3858. doi:10.1063/1.1479463.

[12] J. Nogués, I.K. Schuller, Exchange bias, J. Magn. Magn. Mater. 192 (1999) 203–232. doi:10.1016/S0304-8853(98)00266-2.

[13] C. Schanzer, V.R. Shah, T. Gutberlet, M. Gupta, P. Böni, H.B. Braun, Magnetic depth profiling of FM/AF/FM trilayers by PNR, in: Phys. B Condens. Matter, 2005. doi:10.1016/j.physb.2004.10.044.

[14] L. Xi, Z. Zhang, J.M. Lu, J. Liu, Q.J. Sun, J.J. Zhou, S.H. Ge, F.S. Li, The high-frequency soft magnetic properties of FeCoSi/MnIr/FeCoSi trilayers, Phys. B Condens. Matter. 405 (2010) 682–685. doi:10.1016/j.physb.2009.09.086.

[15] M. Tafur, M.A. Sousa, F. Pelegrini, V.P. Nascimento, E. Baggio-Saitovitch, Ferromagnetic resonance study of dual exchange bias field behavior in NiFe/IrMn/Co trilayers, Appl. Phys. Lett. 102 (2013) 062402. doi:10.1063/1.4791574.

[16] K.W. Lin, T.C. Lan, C. Shueh, E. Skoropata, J. Van Lierop, Modification of the ferromagnetic anisotropy and exchange bias field of NiFe/CoO/Co trilayers through the CoO spacer thicknesses, J. Appl. Phys. 115 (2014) 1–5. doi:10.1063/1.4861216.

[17] J. Moritz, G. Vinai, B. Dieny, Large Exchange Bias Field in ( Pt / Co ) 3 / IrMn / Co Trilayers With Ultrathin IrMn Layers, Ieee Magn. Lett. 3 (2012) 10–13. doi:10.1109/LMAG.2012.2184794.

[18] G. Varvaro, F. Albertini, E. Agostinelli, F. Casoli, D. Fiorani, S. Laureti, P. Lupo, P. Ranzieri, B. Astinchap, A.M. Testa, Magnetization reversal mechanism in perpendicular exchange-coupled Fe/L1 0-FePt bilayers, New J. Phys. 14 (2012). doi:10.1088/1367-2630/14/7/073008.

[19] C. Feng, B.H. Li, J. Teng, Y. Jiang, G.H. Yu, Influence of antiferromagnetic FeMn on magnetic properties of perpendicular magnetic thin films, Thin Solid Films. 517 (2009) 2745–2748. doi:10.1016/j.tsf.2008.10.039.

[20] I. Dzhun, G. Babaytsev, N. Chechenin, Ch. Gritsenko, V. Rodionova, FMR investigations of exchange biased NiFe/IrMn/NiFe trilayers with high and low Ni relative content, Journal of Magnetism and Magnetic Materials (2017), https://doi.org/10.1016/j.jmmm.2017.11.028.

[21] S.L. Gnatchenko, D.N. Merenkov, A.N. Bludov, V. V. Pishko, Y.A. Shakhayeva, M. Baran, R. Szymczak, V.A. Novosad, Asymmetrically shaped hysteresis loop in exchange-biased FeNi/FeMn film, J. Magn. Magn. Mater. 307 (2006) 263–267. doi:10.1016/j.jmmm.2006.04.011.

[22] L.X. Phua, N.N. Phuoc, C.K. Ong, Investigation of the microstructure, magnetic and microwave properties of electrodeposited $Ni_xFe_{1-x}$ (x=0.2–0.76) films, J. Alloys

Compd. 520 (2012) 132–139. doi:10.1016/j.jallcom.2011.12.164.

[23] Z.B. Guo, Y.K. Zheng, K.B. Li, Z.Y. Liu, P. Luo, Y.H. Wu, Asymmetrically kinked hysteresis loops in exchange biased NiFe/IrMn rings, J. Appl. Phys. 95 (2004) 4918–4921. doi:10.1063/1.1690113.

[24] I.L. Castro, V.P. Nascimento, E.C. Passamani, A.Y. Takeuchi, C. Larica, M. Tafur, F. Pelegrini, The role of the (111) texture on the exchange bias and interlayer coupling effects observed in sputtered NiFe/IrMn/Co trilayers, J. Appl. Phys. 113 (2013). doi:10.1063/1.4804671.

[25] G. Malinowski, M. Hehn, S. Robert, O. Lenoble, A. Schuhl, P. Panissod, Magnetic origin of enhanced top exchange biasing in Py/IrMn/Py multilayers, Phys. Rev. B. 68 (2003) 184404. doi:10.1103/PhysRevB.68.184404.

[26] I. Dzhun, N. Chechenin, K. Chichay, V. Rodionova, Dependence of Exchange Bias Field on Thickness of Antiferromagnetic Layer in NiFe/IrMn Structures, Acta Physica Polonica A, 127 (2) (2015) 555-557, DOI: 10.12693/APhysPolA.127.555.

[27] Radu, Florin, and Hartmut Zabel. "Exchange bias effect of ferro-/antiferromagnetic heterostructures." Magnetic heterostructures. Springer, Berlin, Heidelberg, 2008. 97-184.

[28] Xi, Haiwen, and Robert M. White. "Antiferromagnetic thickness dependence of exchange biasing." Physical Review B 61.1 (2000): 80.

[29] Sankaranarayanan, V. K., et al. "Exchange bias in Ni Fe∕ Fe Mn∕ Ni Fe trilayers." Journal of applied physics 96.12 (2004): 7428-7434.

[30] Nascimento, V. P., et al. "Influence of the roughness on the exchange bias effect of NiFe/FeMn/NiFe trilayers." Journal of Magnetism and Magnetic Materials 320.14 (2008): e272-e274.

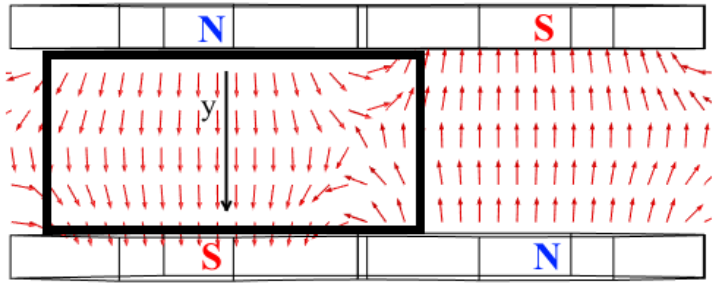

**Figure 1.** Visualization of magnetic field between permanent magnets in the location of substrate (the black frame) during the deposition simulated by *Comsol Multiphysics*.

| 404 | 576 | 587 | 571 | 567 | 573 | 604 | 655 | 558 | -259 |
|-----|-----|-----|-----|-----|-----|-----|-----|-----|------|
| 385 | 460 | 500 | 513 | 518 | 511 | 457 | 302 | 266 | -37  |
| 257 | 355 | 435 | 468 | 488 | 485 | 456 | 369 | 204 | -52  |
| 236 | 350 | 440 | 486 | 504 | 503 | 486 | 413 | 254 | -38  |
| 286 | 515 | 597 | 585 | 573 | 573 | 588 | 630 | 581 | -170 |

**Figure 2.** Map of the magnetic field magnitude between the two permanent magnets in the region of substrate, measured by a Hall probe. The dashed lines correspond to the borders of each region (values are given in Oe). The large numbers indicate the regions.

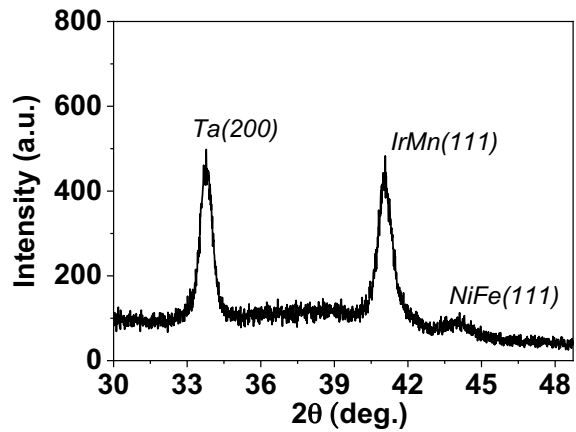

**Figure 3.** XRD pattern for Ta(30 nm)/Ni$_{75}$Fe$_{25}$(10 nm)/Ir$_{45}$Mn$_{55}$(20 nm)/Ta(30 nm) structure.

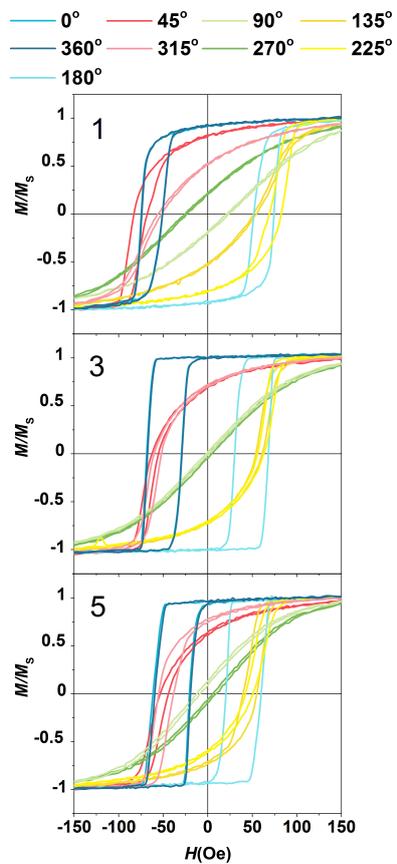

**Figure 4.** Hysteresis loops for regions 1, 3, 5, measured at different azimuthal angles. The numbers indicate the regions.

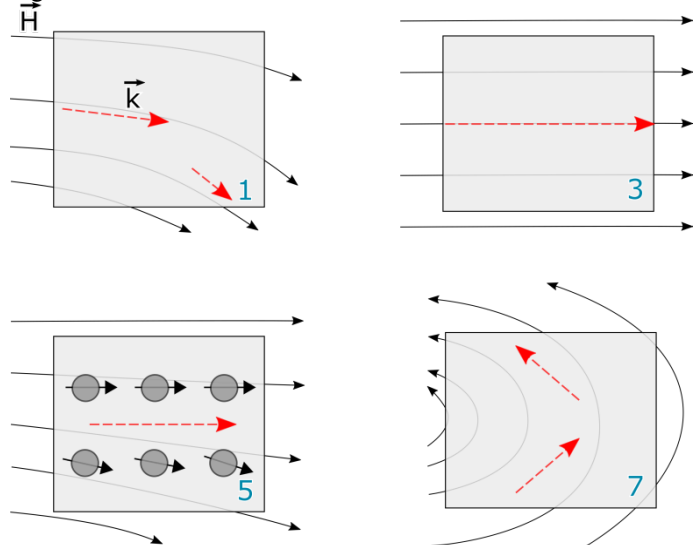

**Figure 5.** Scheme of an inhomogeneous magnetic field influence on the formation of unidirectional anisotropy, k. The magnetic field lines are shown by black color, the blue numbers indicate the regions.

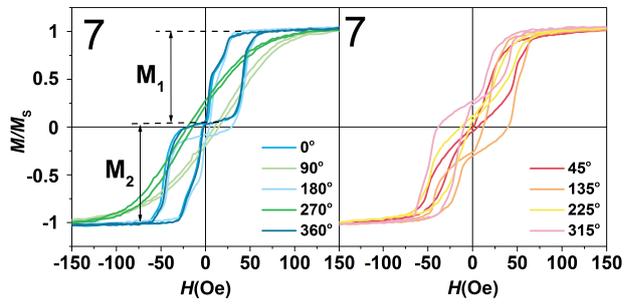

**Figure 6.** The hysteresis loops for region 7 measured at different azimuthal angles.

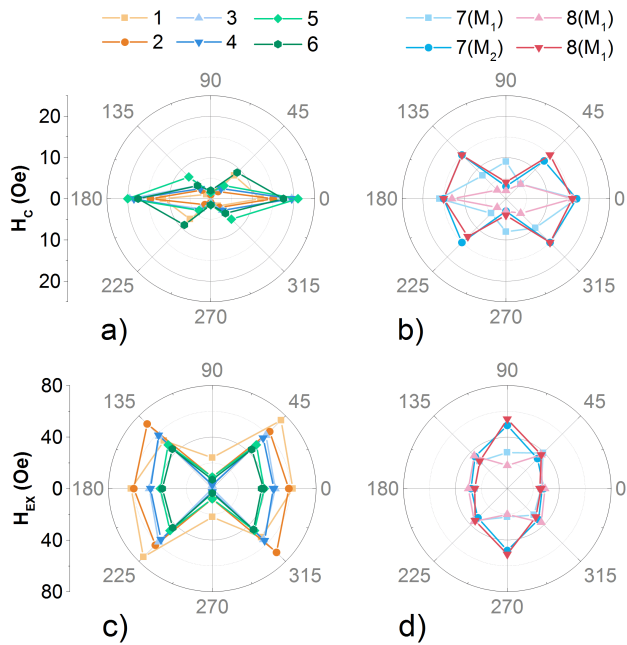

**Figure 7.** The angular dependences of exchange bias a), c) and coercive forces b), d). The numbers indicate the regions, whereas $M_1$ and $M_2$ correspond to the sub-loops.